\documentclass[aps,prb,twocolumn,superscriptaddress]{revtex4}
\usepackage{graphicx}
\usepackage{amsmath}
\usepackage{amssymb}
\usepackage{comment}

\begin{document}

\title{Evaporative depolarization and spin transport in a unitary trapped Fermi gas}

\author{Meera M. Parish}
\affiliation{Department of Physics, Princeton University, Princeton,
NJ 08544} %
\affiliation{Princeton Center for Theoretical Science, Princeton
University, Princeton, NJ 08544} %

\author{David A. Huse}
\affiliation{Department of Physics, Princeton University, Princeton,
NJ 08544} %

\date{
\today}

\begin{abstract}
We consider a partially spin-polarized atomic Fermi gas in a
high-aspect-ratio trap, with a flux of predominantly spin-up atoms
exiting the center of the trap. We argue that such a scenario can be
produced by evaporative cooling, and we find that it can result in a
substantially non-equilibrium polarization pattern for typical
experimental parameters. We offer this as a possible explanation for
the quantitative discrepancies in recent experiments on
spin-imbalanced unitary Fermi gases.
\end{abstract}



\maketitle

\section{Introduction}
Two-component atomic Fermi gases provide an ideal experimental
system in which to investigate fermion pairing and superfluidity in
a controllable manner.~\cite{giorgini2008} For example, one can use
a magnetically-tunable Feshbach resonance to access the unitary
regime, where the scattering length diverges and one has a
strongly-interacting fermionic superfluid that is
`universal'.~\cite{ho2004}
Of particular interest is the case where there is a spin imbalance
that frustrates pairing between fermion species, because this is a
situation that arises in many fields of physics, ranging from QCD to
superconductivity.~\cite{casalbuoni2004}
%
Here, a central question has been: what is the critical spin
polarization $\delta_c$ at which pairing and superfluidity are
destroyed for a unitary trapped Fermi gas at equilibrium?
However, current experiments produce different answers. The
experimental group at MIT finds that $\delta_c \simeq
77\%$,~\cite{zwierlein2006,zwierlein2006_2,shin2006,shin2008} while
experiments on highly-elongated trapped gases at Rice
University~\cite{partridge2006_2} suggest that $\delta_c$ is at
least 90\%.
Moreover, even though the critical polarization is known to be a
strong function of temperature, where $\delta_c$ decreases with
increasing temperature, the different $\delta_c$'s observed in
experiment are unlikely to be caused by differences in temperature
since both experimental groups have claimed that their temperatures
are low enough to yield $\delta_c$'s that are close to the
zero-temperature result.
Thus, there is a real discrepancy in the measured $\delta_c$ and a
resolution of this problem has potentially important implications
for the nature of the paired superfluid phase in a finite-sized
system.
Here we propose that the high $\delta_c$ observed in the Rice
experiment was due to their trapped spin-imbalanced gas being out of
equilibrium. We show that a combination of the trap geometry and the
evaporative cooling scheme implemented in the Rice experiment can
induce a spin current along the trap axis, which in turn creates a
substantially non-equilibrium polarization pattern that favors a
superfluid at the trap center.

\begin{figure}
\centering
\includegraphics[width = 0.47\textwidth]{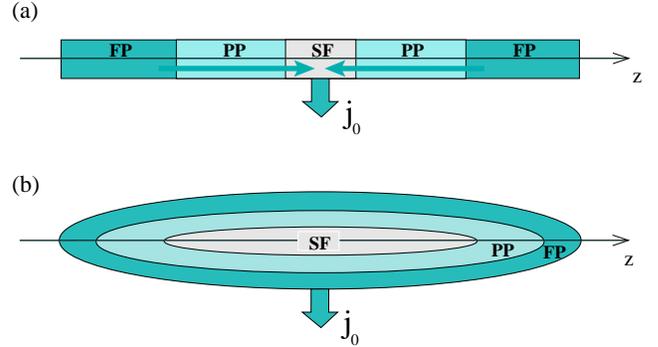}
\caption{(Color online) Schematic diagram of atom transport during
the evaporative cooling process in trapped, spin-imbalanced,
highly-degenerate Fermi gases in the quasi-1D (a) and 3D (b)
regimes. The flux of spin-up atoms $j_0$ is exiting the gas over the
lip of the trap that is located at the axial center ($z=0$).  This
flux must be drawn from the fully-polarized (FP) normal region.
Thus, for a quasi-1D gas (a), a spin current must flow through the
intervening partially-polarized (PP) normal region
and across the normal/superfluid(SF) interface. In the 3D regime
(b), the atoms can evaporate directly from the surrounding
fully-polarized layer.} \label{fig:spin_current}
\end{figure}

To understand how such a spin current 
can be generated, one must first examine the evaporative cooling
process.  Here, the temperature, and entropy per atom, of a trapped
gas is lowered when the most energetic atoms escape over the ``lip''
of the trap  --- the route of escape with the lowest potential
barrier to be surmounted.
For a partially-polarized Fermi gas at temperature $T$, the rate of
thermal activation over this barrier is larger for the majority
species by a factor of $\exp{[(\mu_\uparrow-\mu_\downarrow)/k_BT]}$,
assuming the two species are subject to the same trapping potential,
where $\mu_\uparrow$, $\mu_\downarrow$ are the chemical potentials
for the majority and minority species, respectively. Thus, at low T,
the flux of evaporating atoms passing over the lip is essentially
fully polarized, and we have \textit{evaporative depolarization} in
addition to evaporative cooling (as stated in
Ref.~\onlinecite{zwierlein2006}).
%
The Rice experiments we are considering~\cite{partridge2006,
partridge2006_2} had both a long, thin, high-aspect-ratio optical
trap and a nonuniform magnetic field that contributed to the axial
confinement of the gas.
As a result, the lowest barrier for the atoms to escape from the
trap was at the axial center, with the atoms escaping over this
``lip'' in the radial direction (and downwards, due to gravity; see
Fig.~\ref{fig:spin_current}(a)).  Elsewhere in the trap the barrier
to escape was significantly higher, so at their lowest temperatures,
essentially all of the evaporating atoms escaped radially at the axial
center of their high-aspect-ratio trap.

To achieve low temperatures in the Rice experiment, the height of
this barrier was lowered by reducing the intensity of the optical
trap.
It was at the lowest temperatures that the unusually large 
$\delta_c$ was observed, along with strong deviations from the
equilibrium local density approximation (LDA) in the shapes of the
regions occupied by the superfluid and partially-polarized normal
phases.~\cite{partridge2006_2}
What we propose happened here is that the evaporation, with the
$\uparrow$ atoms rapidly escaping radially over the trap lip,
greatly depleted any excess unpaired $\uparrow$ atoms from the
axially central region of the cloud occupied by the paired
superfluid phase. This depletion, which is apparent in the
\textit{in situ} density measurements,~\cite{partridge2006,
partridge2006_2} substantially suppressed
$(\mu_{\uparrow}-\mu_{\downarrow})$ in that region (evaporative
depolarization).  The flux of evaporating $\uparrow$ atoms over the
lip then had to come from the fully-polarized normal regions at the
axial ends of the cloud and be
driven through the partially-polarized region and across the 
normal/superfluid interface by a substantial axial gradient of
$(\mu_{\uparrow}-\mu_{\downarrow})$.
This resulted in the partially-polarized region of the cloud being
much smaller in axial extent than it would be at equilibrium, which
is the strong deviation from equilibrium LDA that we will focus on
in this paper.
There was another important deviation from equilibrium LDA: the
aspect ratio of the central superfluid region was substantially
reduced from that of the cloud as a whole; it is this latter feature
that was emphasized in Ref.~\onlinecite{partridge2006_2}.

We emphasize that the non-equilibrium scenario described above
assumes that the atom transport is effectively one-dimensional (1D)
along the axial ($z$) direction, as illustrated in
Fig.~\ref{fig:spin_current}(a).~\cite{1Dnote} Thus, it is only
appropriate for high-aspect-ratio gas clouds with sufficiently low
particle number, like those in the Rice experiments (aspect ratio
$\gtrsim30:1$) at the lowest temperatures. For higher temperatures
(less evaporative cooling), where the particle numbers are larger,
there is a fully-polarized layer of $\uparrow$ atoms \textit{fully}
surrounding the cloud (Fig.~\ref{fig:spin_current}(b)), as can be
seen in \textit{in situ} density
measurements.~\cite{partridge2006_2} In this case, the evaporation
will simply draw $\uparrow$ atoms from this fully-polarized layer
rather than driving a spin current through the superfluid and
partially-polarized normal regions, so a strong chemical potential
gradient is not produced.
The MIT experiments, on the other hand, used trapped gases with a
much lower aspect ratio ($\sim5:1$) and at least an order of
magnitude more atoms in the cloud.  Thus, the 3D scenario for
highly-polarized gases depicted in Fig.~\ref{fig:spin_current}(b)
remains correct for these experiments, even at the lowest
temperatures, and so their observed $\delta_c$ agrees with the
equilibrium result from quantum Monte Carlo (QMC)
calculations.~\cite{lobo2006}

\section{Model of evaporation}
To model the spin transport in the Rice experiment at the lowest
temperatures, we will assume that atoms are only removed from the
gas at $z=0$, the position of the trap lip, and that only $\uparrow$
atoms are evaporating (Fig.~\ref{fig:spin_current}(a)).
Also, we approximate any nonuniformities in the local chemical potentials
as 1D, so that we only consider non-equilibrium
\emph{axial} gradients in $\mu_\uparrow$, $\mu_\downarrow$.
%
Since we are primarily interested in the non-equilibrium
polarization pattern within a trap, and this is not strongly
dependent on the local temperature, we will ignore any gradients in
temperature.
We further assume that the atomic gas cloud is in \emph{mechanical}
equilibrium throughout the evaporation process,
since the time required to equalize a pressure difference 
in the cloud is much smaller than both the duration of the
evaporation and the time for spin diffusion along the length of the
cloud. This implies that the local pressure $P$ always satisfies:
\begin{align} \label{eq:mecheq}
\frac{\partial P}{\partial z} = & -n \frac{\partial V}{\partial z}
\; ,
\end{align}
where $n \equiv n_\uparrow + n_\downarrow$ is the \emph{local} total
density and $V(z)$ is the axial trapping potential.

\section{Spin transport in a unitary normal gas}
We can determine $\delta_c$ of a spin-imbalanced Fermi gas 
in the presence of a spin current by focussing on the unitary
\textit{normal} gas at low temperatures.
Here, the equation of state for the pressure at $T=0$ is accurately
known from QMC calculations~\cite{lobo2006,pilati2008} and is given
by
\begin{align}\label{eq:pressure}
 P_N = & \frac{2}{5} n_\uparrow \varepsilon_{F\uparrow} \left[ 1 - A
 \frac{n_\downarrow}{n_\uparrow} + \frac{m}{m^*}\left(\frac{n_\downarrow}{n_\uparrow}
 \right)^{5/3} + F \left(\frac{n_\downarrow}{n_\uparrow}\right)^2 \right]
\end{align}
where $m$ is the atomic mass, $m^* \simeq 1.09 m \simeq m$, $A
\simeq 0.99$ and $F\simeq 0.14$.
For the spin transport that we are interested in, we also expect
Eq.~\eqref{eq:pressure} to provide a reasonable approximation for
the pressure at low temperatures within the degenerate regime $T <
\varepsilon_{F\uparrow}/k_B \equiv T_{F\uparrow}$, where
$\varepsilon_{F,\sigma} = \hbar^2(6\pi^2 n_\sigma)^{2/3}/2m$.
%
%
%
%
%
In addition to Eq.~\eqref{eq:mecheq}, we require equations for the
\emph{spin} density $n_s = n_\uparrow - n_\downarrow$ and
\emph{spin} current density $j_s = j_\uparrow - j_\downarrow$. We
will assume that the ``DC'' spin transport in the
partially-polarized normal Fermi gas is diffusive, with the
dissipation being driven by the interspecies interactions. Indeed,
this situation may be regarded as the cold-atom analogue of spin
Coulomb drag in electron systems.~\cite{spin_drag}
By transforming to the inertial reference frame where locally
$j^\prime = j^\prime_\uparrow + j^\prime_\downarrow = 0$, i.e.\
where there is no net transport of the total mass density and the
motion is purely diffusive, we obtain current densities:
\begin{align} \label{eq:frame_dens}
j^\prime_\sigma = & -D_\sigma \left(\frac{\partial
n_\sigma}{\partial z} - \frac{\partial n_\sigma^{eq}}{\partial z}
\right)~,
\end{align}
where $\frac{\partial n_\sigma^{eq}}{\partial z}$ is the local
equilibrium density gradient and $D_\sigma$ is the diffusion
constant for each spin. One can determine $\frac{\partial
n_\sigma^{eq}}{\partial z}$ as a function of the local densities
using the LDA equilibrium condition for the chemical potentials of
each spin:
\begin{align} 
\frac{\partial \mu_\sigma}{\partial z} = - \frac{\partial
V}{\partial z}~.
\end{align}
Note that this automatically implies that there are no gradients in
the chemical potential difference $(\mu_\uparrow - \mu_\downarrow)$
at equilibrium.
%
%
%
Combining Eq.~\eqref{eq:frame_dens} with the expression for the spin
current density in this frame
\begin{align}
j^\prime_s = & j^\prime_\uparrow - j^\prime_\downarrow \equiv -D_s
\left(\frac{\partial n_s}{\partial z} - \frac{\partial
n_s^{eq}}{\partial z} \right)~,
\end{align}
yields an expression for the spin diffusion constant:
\begin{align}
\frac{1}{D_s} = & \frac{1}{2} \left(\frac{1}{D_\uparrow} +
\frac{1}{D_\downarrow} \right)~.
\end{align}
Transforming back to the lab reference frame then gives us the
spin current density:
\begin{align} \label{eq:spincurr}
 j_s & 
 = -D_s \left(\frac{\partial n_s}{\partial z} - \frac{\partial
n_s^{eq}}{\partial z} \right) + n_s v~,
\end{align}
where the average net velocity $v = j/n$.

For a dilute gas with $s$-wave scattering, we have diffusion
constants $D_\sigma = \frac{1}{3} v_\sigma^2 \tau_\sigma$, where
$v_\sigma$ is the average velocity of the random particle motion and
$1/\tau_\sigma$ is the scattering rate of each species. In the
regime $T\lesssim T_{F\downarrow} < T_{F\uparrow}$, the scattering
rate for the $\downarrow$ atoms at unitarity is:~\cite{bruun2008}
\begin{align}
 \frac{1}{\tau_\downarrow} \simeq & \frac{4 \pi^3 A^2}{25}
 \frac{m^* (k_B T)^2 }{\hbar^3 k_{F\uparrow}^2}~,
\end{align}
By knowing $\tau_{\downarrow}$ and $v_\sigma$, we can also extract
an expression for the $\uparrow$ scattering rate since the mean free
paths $l_\sigma = v_\sigma \tau_\sigma$ are simply related by the
densities: $l_\uparrow = l_\downarrow (n_\uparrow/n_\downarrow)$.
One can see this from the fact that $l_\sigma =
1/\sigma_{cs}n_{-\sigma}$ in a dilute gas, where the scattering
cross section $\sigma_{cs}$ is a universal function of the density
at unitarity. Also, we can approximate the velocities as $v_\sigma
\simeq \hbar k_{F\sigma}/m$.
Thus, we obtain the `universal' spin diffusion constant:
\begin{align} \label{eq:diffusT}
D_s \simeq & \frac{50}{3\pi^3 A^2
}\frac{\hbar}{m}\left(\frac{T_{F\downarrow} T_{F\uparrow}}{T^2}
\right)\left[1 + \left( \frac{n_\downarrow}{n_\uparrow}
\right)^{4/3}\right]^{-1}~.
\end{align}
%
%
Note that the dependence of $D_s$ on temperature is that of a Fermi
liquid, as expected.
Experiments to explicitly measure this `universal' transport
constant would be welcome.

Referring to Fig.~\ref{fig:spin_current}(a), if the fully-polarized
normal regions at the axial ends are sufficiently large (i.e.\ if
the gas has a sufficiently high global polarization), then we can
treat them as stationary spin current sources and approximate the
spin current through the partially-polarized region as
\emph{steady-state}:
\begin{align} 
\frac{\partial n_\sigma}{\partial t}= -\frac{\partial
j_\sigma}{\partial z} \simeq 0~.
\end{align}
In this case, by solving Eq.~\eqref{eq:spincurr} with $j_s = j =
j_0$, together with Eq.~\eqref{eq:mecheq},
we find that the deviation of the
\textit{total} polarization from equilibrium for a given
\textit{local} polarization at the trap center is determined by the
simple dimensionless quantity:
\begin{align}
 Q \equiv & \frac{|j_0|L}{n_{\uparrow}(z=0)}
\left(\frac{T}{T_{F\uparrow}(z=0)}\right)^2 \frac{m}{\hbar}~,
\end{align}
where 
$L$ is the total axial length of the cloud.

The critical total polarization $\delta_c$ for a trapped cloud is
obtained by setting the local densities $n_\downarrow/n_\uparrow
\cong 0.44$ at the trap center, which is its value at the
superfluid-normal transition in the uniform system.~\cite{lobo2006}
This corresponds to the situation where a trapped normal gas is on
the verge of forming a superfluid core.
To obtain a quantitative estimate of $\delta_c$ as a function of $Q$
in a high-aspect-ratio harmonic trap, we determine the chemical
potentials as a function of $z$ and then we use LDA in the radial
direction $r$ to now include the radial trapping potential $V_\perp(r)$
and obtain the densities of each spin:
\begin{align}
n_\sigma(z,r)  \equiv n_\sigma\left[\mu_{\uparrow}(z) -
V_\perp(r),\mu_{\downarrow}(z)- V_\perp(r)\right]~.
\end{align}
Integrating these densities numerically then yields the total
polarization $\delta \equiv (N_\uparrow - N_\downarrow)/(N_\uparrow
+ N_\downarrow)$.

\begin{figure}
\centering
\includegraphics[width = 0.47\textwidth]{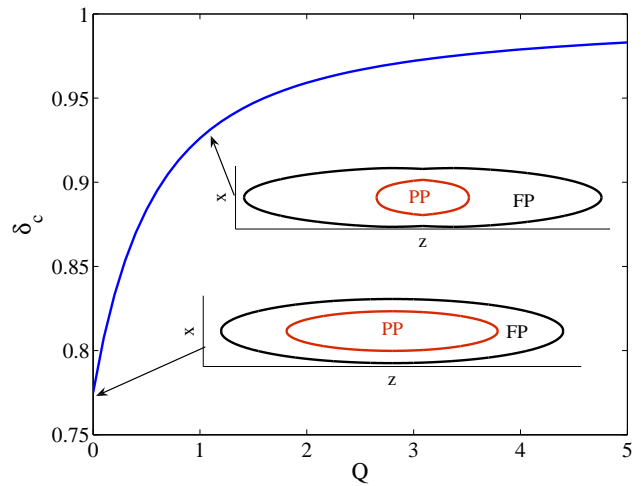}
\caption{(Color online) Total critical polarization $\delta_c$ of a
trapped quasi-1D spin-imbalanced Fermi gas as a function of the
dimensionless spin current $Q$. The figure insets depict cross
sections of the trapped normal gas at $\delta_c$ for $Q=0$ and
$Q=1$. In the latter case, the partially-polarized (PP) normal core
has been distorted and shrunk compared to the fully-polarized (FP)
region, in clear violation of LDA. Note that the depicted trap
aspect ratio is at least 5 times smaller than that in the Rice
experiment, and thus the apparent FP layer surrounding the PP core
near the axial center ($z=0$) is extremely thin, being only a couple
of atoms thick, and does not violate our approximation of 1D
transport.} \label{fig:critpol}
\end{figure}

As depicted in Fig.~\ref{fig:critpol}, the critical polarization $\delta_c$
corresponds (by construction) to the equilibrium QMC
result~\cite{lobo2006} when $Q = 0$, but it dramatically increases
with increasing $Q$ and asymptotically approaches 100\% polarization
as $Q\rightarrow \infty$.
%
This increase in $\delta_c$ is due to the partially-polarized region
being 
compressed along the axial direction relative to its extent in
equilibrium LDA (inset of Fig.~\ref{fig:critpol}), an effect which
has been observed in the Rice experiments.
To obtain $\delta_c \simeq 90 \%$, we require an evaporation rate of
order $10^6$ atoms/sec if we use typical parameter values in the
Rice experiment towards the end of evaporation: $T/T^0_{F\uparrow}
\simeq 0.05$, $L \simeq 1$mm, $n^0_{\uparrow} \simeq
10^{12}$cm$^{-3}$, $\hbar/m \simeq 10^4\mu$m$^2$/s, and a trap
radius of 10 $\mu$m.  This appears consistent with the data in
Ref.~\onlinecite{partridge2006_2} since the actual evaporation rate
is estimated to be roughly 10$^6$ atoms/sec at the time when the
optical absorption images are taken.~\cite{partridge}

\section{Superfluid/normal interface}
\begin{figure}
\centering
\includegraphics[width = 0.35\textwidth]{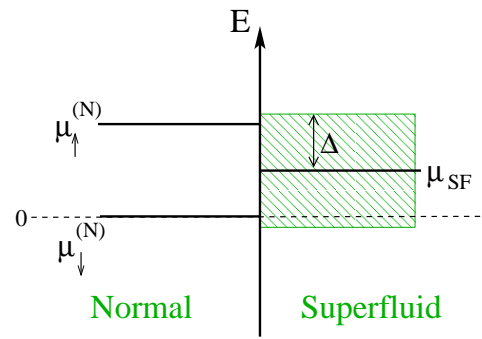}
\caption{(Color online) Interface between the unpolarized superfluid
and partially-polarized normal state at unitarity, where the local
chemical potentials are shown for the interface at local
equilibrium, with superfluid chemical potential $2\mu_{SF} \equiv
\mu_{\uparrow}^{(SF)} + \mu_{\downarrow}^{(SF)}$. The shaded energy
range corresponds to the quasiparticle gap (the spin gap) of the
superfluid, with $\Delta \simeq 1.2 \mu_{SF}$ taken from $T=0$ QMC
calculations.~\cite{carlson2008}}\label{fig:interface}
\end{figure}

For completeness, we now examine the effect of a superfluid core on
the spin current in the trapped gas when $\delta < \delta_c$. For
this purpose, we consider spin transport through the
superfluid/normal interface. We approximate the interface to be a 1D
sharp step-function in the local superfluid gap $\Delta(z)$ and
densities (Fig.~\ref{fig:interface}), which is reasonable when $T$
is well below the temperature at the tricritical
point.~\cite{parish2007} Clearly, our assumption of mechanical
equilibrium implies that the pressures in each phase are equal at
the interface: $P_N = P_{SF}$.
Since we are focussing on low $T$, we will use the $T=0$ expression
for the superfluid pressure:
\begin{align}
  P_{SF} = \frac{2}{15\pi^2} \left(\frac{2m}{\xi \hbar^2}
\right)^{3/2} \mu_{SF}^{5/2} ~,
\end{align}
where $\xi \simeq 0.42$ according to QMC
calculations,~\cite{carlson2005} and the superfluid chemical
potential $2\mu_{SF} \equiv \mu_{\uparrow}^{(SF)} +
\mu_{\downarrow}^{(SF)}$.
We further assume that any quasiparticles in the superfluid region
are rapidly evaporated so that we can neglect any quasiparticles
incident on the interface from the superfluid side. Effectively,
this amounts to assuming a strong drop in the local temperature of
the quasiparticles as one crosses the interface from the normal to
the superfluid side. While this is not essential to our picture,
this gives a simplification and likely captures what happens when
the lip of the trap is held very low.


The scattering problem at the superfluid/normal interface has
already been examined in Ref.~\onlinecite{schaeybroeck2007}.
However, they only focussed on thermal transport via quasiparticle
transmission through the interface. In our case, we must consider
both mass and spin transport across the interface, since they can
both change the local polarization. We must therefore take account
of any mass transported via Andreev reflection.
For example, if the chemical potentials $\mu_\uparrow^{(N)}$,
$\mu_\downarrow^{(N)}$ in the normal phase lie within the gapped
region of the superfluid, like in Fig.~\ref{fig:interface}, the
Andreev process results in a flux of pairs (mass) into the
superfluid when $2\mu_{SF} < \mu_{\uparrow}^{(N)} +
\mu_{\downarrow}^{(N)}$ and a flux out of the superfluid when
$2\mu_{SF} > \mu_\uparrow^{(N)} + \mu_\downarrow^{(N)}$, even at
$T=0$.

Following the approach of Ref.~\onlinecite{BTK82}, we can write the
current density of $\uparrow$ atoms flowing into the superfluid as:
\begin{align} \notag
j_\uparrow = & \frac{1}{(2\pi)^2\hbar} \int dE \int dk_\uparrow
\left[k_\uparrow\left(1 - B_\uparrow\right)
f(E-\mu_\uparrow^{(N)}) \right. \\ \label{eq:spin_up}
& \left.
- k_\downarrow A_\uparrow
f(E-2\mu_{SF}+\mu_\downarrow^{(N)}) \frac{}{} \right]
\end{align}
where $k_\uparrow$ ($k_\downarrow$) is the momentum normal to the
interface of the incident $\uparrow$ atom (reflected $\downarrow$
hole), $E$ is the energy and $f(x)$ is the Fermi function.
The coefficients $A_\uparrow$ and $B_\uparrow$ correspond to the
probability of Andreev reflection and ordinary reflection,
respectively. Clearly, when $|E - \mu_{SF}| < \Delta$, there is no
quasiparticle transmission and we must have $k_\uparrow = k_\uparrow
B_\uparrow + k_\downarrow A_\uparrow$.
%
Also, for $E < \Delta + \mu_{SF}$ (appropriate for low $T$), the
current density of \textit{$\downarrow$ holes} crossing the
interface into the superfluid has the same expression as in
Eq.~\eqref{eq:spin_up}, but with $\uparrow$ and $\downarrow$
interchanged, and $f(x)\rightarrow f(-x)$.
Note that although our calculation of reflection probabilities
relies on the mean-field Bogoliubov-de-Gennes equations, the
chemical potentials, pressure and superfluid gap that we use are
from $T=0$ QMC results.~\cite{pilati2008,carlson2005,carlson2008}

Referring to Fig.~\ref{fig:interface}, if we consider chemical
equilibrium, where $2\mu_{SF} = \mu_\uparrow^{(N)} +
\mu_\downarrow^{(N)}$, then at finite $T$ we find that we have both
a flux of $\uparrow$ atoms entering the superfluid region and a flux
of $\downarrow$ atoms leaving the superfluid region (or a flux of
$\downarrow$ holes flowing in the opposite direction). However,
atoms, not holes, are removed during evaporative cooling and thus we
require the flux $j_\downarrow = 0$ across the interface. This is
achieved by setting $2\mu_{SF} \gtrsim \mu_\uparrow^{(N)} +
\mu_\downarrow^{(N)}$ such that the concomitant mass current flowing
into the superfluid cancels the $\downarrow$ hole current, so there
is only a net motion of $\uparrow$ atoms.
Thus, the superfluid core ensures that the flux of
evaporating atoms remains spin-polarized even when $(\mu_\uparrow -
\mu_\downarrow)$ is suppressed at the trap center due to local
evaporative depolarization.

\section{Concluding remarks}
%
To summarize, we have shown that evaporative cooling of a trapped,
partially-polarized, high-aspect-ratio atomic Fermi gas can create a
spin current which in turn can produce the large $\delta_c$ seen in
the Rice experiment.
Moreover, when this low-temperature gas cloud develops a superfluid
core, we find that the atom flux across the superfluid-normal
interface remains spin-polarized, with a small drop in the average
chemical potential across the interface.
An alternative \textit{equilibrium} explanation for the large
$\delta_c$ in the Rice experiment is that finite-size effects in a
high-aspect-ratio trapped gas are
responsible.~\cite{sensarma2007,ueda,ku2009} However, microscopic
studies of the surface tension at the superfluid/normal interface in
a trapped gas suggest that this scenario is not quantitatively
consistent with experiments.~\cite{baur2009}
Two possible avenues for the experiments to differentiate between
equilibrium and non-equilibrium scenarios are: (1) To explicitly
look for relaxation (on time scales that we estimate to be $\lesssim
0.1$ sec) towards equilibrium after deepening the optical trap and
thus slowing or stopping the evaporative depolarization; and (2) To
move the trap lip off-center with respect to the optical trap
minimum by moving the minimum of the magnetic potential. If our
non-equilibrium explanation is correct, at the lowest temperatures
the superfluid core should form where the local evaporative
depolarization is occurring at the trap lip; at equilibrium the
superfluid will instead form at the minimum of the overall trapping
potential where the density is highest. Indeed, a very recent
experiment~\cite{salomon2009} by Salomon's group at the ENS already
casts some doubt on the alternative equilibrium explanation: their
trap aspect ratio ($\sim 22:1$) approaches that used in the Rice
experiment, yet they observe a $\delta_c$ that agrees with the MIT
result.
Whilst it is still possible that this lower $\delta_c$ may be
because the temperature in the ENS experiment is higher, another key
difference between the Rice and ENS experiments is that the time
scale for evaporation in the ENS experiment is longer,~\cite{hulet}
a fact which appears to be consistent with our non-equilibrium
proposal.

More generally, the non-equilibrium scenario we have analyzed here
suggests a route towards the realization and investigation of spin
currents in a strongly-interacting Fermi gas.
Moreover, it illustrates the importance of non-equilibrium phenomena
in cold-atom systems.
In typical experiments with ultracold atoms, the dynamic range
between the microscopic time scales for atom motion, atom-atom
interactions and the total duration of the experiments is not so
large, often 5 orders of magnitude or less.  This means that full
equilibration of a large cloud of atoms might not be possible under
some circumstances.  Here we have discussed a trapped unitary Fermi
gas, where the microscopic collision rate at low temperature is
roughly $\varepsilon_F/\hbar$.  When one instead considers atoms in
an optical lattice, the microscopic rates for atom hopping or for
superexchange can be substantially slower than this, making it even
easier to drive the system out of equilibrium.~\cite{chin2009}

\acknowledgments We thank R. Hulet, W. Li, G. Partridge, W. Ketterle
and M. Zwierlein for discussions and information about their
experiments. This work is supported under ARO Award W911NF-07-1-0464
with funds from the DARPA OLE Program.


\end{document}